\documentclass[sigconf,pbalance,nonacm]{acmart}

\AtBeginDocument{%
  }

\setcopyright{none}

\usepackage{pifont}
\usepackage{xcolor}
\usepackage{cuted}
\usepackage{subcaption}
\usepackage{listings}
\usepackage[frozencache]{minted}
\usepackage{multirow}
\usepackage{booktabs}

\DeclareUnicodeCharacter{2265}{\ensuremath{\ge}}

\usepackage{array}
\usepackage[para,online,flushleft]{threeparttable}

\newcommand{\cmark}{\textcolor{green!70!black}{\ding{51}}} 
\newcommand{\xmark}{\textcolor{red!70!black}{\ding{53}}}   

\def\clap#1{\hbox to0pt{\hss#1\hss}}

\usepackage{tikz}
\usetikzlibrary{positioning, fit, backgrounds}

\begin{document}

\title{A High-level Synthesis Toolchain for the Julia Language}

\author{Benedict Short}
\affiliation{%
  \institution{Imperial College London}
  \city{London}
  \country{UK}
}

\author{Ian McInerney}
\email{i.mcinerney17@imperial.ac.uk}
\orcid{0000-0003-2616-9771}
\affiliation{%
  \institution{Imperial College London}
  \city{London}
  \country{UK}}

\author{John Wickerson}
\email{j.wickerson@imperial.ac.uk}
\orcid{0000-0001-6735-5533}
\affiliation{%
  \institution{Imperial College London}
  \city{London}
  \country{UK}
}


\begin{abstract}
With the push towards Exascale computing and data-driven methods, problem sizes have increased dramatically, increasing the computational requirements of the underlying algorithms.
This has led to a push to offload computations to general purpose hardware accelerators such as GPUs and TPUs, and a renewed interest in designing problem-specific accelerators using FPGAs.
However, the development process of these problem-specific accelerators currently suffers from the ``two-language problem'': algorithms are developed in one (usually higher-level) language, but the kernels are implemented in another language at a completely different level of abstraction and requiring fundamentally different expertise.
To address this problem, we propose a new MLIR-based compiler toolchain that unifies the development process by automatically compiling kernels written in the Julia programming language into SystemVerilog without the need for any additional directives or language customisations.
Our toolchain supports both dynamic and static scheduling, directly integrates with the AXI4-Stream protocol to interface with subsystems like on- and off-chip memory, and generates vendor-agnostic RTL.
This prototype toolchain is able to synthesize a set of signal processing/mathematical benchmarks that can operate at 100MHz on real FPGA devices, achieving between 59.71\% and 82.6\% of the throughput of designs generated by state-of-the-art toolchains that only compile from low-level languages like C or C++.
Overall, this toolchain allows domain experts to write compute kernels in Julia as they normally would, and then retarget them to an FPGA without additional pragmas or modifications.
\end{abstract}

\ccsdesc[500]{Hardware~High-level and register-transfer level synthesis}
\ccsdesc[500]{Hardware~Hardware description languages and compilation}
\ccsdesc[500]{Hardware~Hardware accelerators}

\keywords{Julia language, MLIR, High-level Synthesis, CIRCT}

\definecolor{colorNEW}{HTML}{66a61e}
\definecolor{colorMODIFIED}{HTML}{e6ab02}
\definecolor{colorASIS}{HTML}{bdbdbd}

\definecolor{colorFRONTENDfg}{HTML}{ffffff}
\definecolor{colorFRONTENDbg}{HTML}{fbb4ae}

\definecolor{colorBACKENDfg}{HTML}{ffffff}
\definecolor{colorBACKENDbg}{HTML}{b3cde3}

\tikzset{
  pass/.style={draw, fill=white, thick, anchor=west, align=left, minimum width=40mm},
  arr/.style={-latex, thick},
  frontend/.style={draw=none, fill=colorFRONTENDbg, rounded corners},
  backend/.style={draw=none, fill=colorBACKENDbg, rounded corners},
  NEW/.style={draw=none, fill=colorNEW, inner sep=1pt, outer sep=0pt, minimum width=1.5em},
  MODIFIED/.style={draw=none, fill=colorMODIFIED, inner sep=1pt, outer sep=0pt, minimum width=3.1em},
  ASIS/.style={draw=none, fill=colorASIS, inner sep=1pt, outer sep=0pt, minimum width=3.1em},
}

\newcommand\NEW[1]{%
\protect\tikz\protect%
\node[NEW](a){%
\clap{\textcolor{white}{\tiny\sf\bfseries NEW}}%
};%
}

\newcommand\MODIFIED[1]{%
\protect\tikz\protect%
\node[MODIFIED](a){%
\clap{\textcolor{white}{\tiny\sf\bfseries MODIFIED}}%
};%
}

\newcommand\ASIS[1]{%
\protect\tikz\protect%
\node[ASIS](a){%
\clap{\textcolor{white}{\tiny\sf\bfseries USED AS-IS}}%
};%
}

\newcommand\defaultgap{8mm}
\newcommand\arrshift{-13mm}

\maketitle

\section{Introduction}

The end of Moore's Law has meant that scientific algorithms can no longer rely on advancements in computer architecture and fabrication processes for uplifts in CPU performance, and instead must turn to hardware-software codesign methods to both optimize software for the current hardware and potentially design hardware optimized for the application \cite{ChaNei:20}.
Optimizing the software for the current hardware has surged in popularity, with large-scale projects such as the US Exascale Computing Project (ECP) demonstrating its potential for large-scale impact \cite{germannCodesignExascaleComputing2021, herouxECPLibrariesTools2024}.
However, optimizing the hardware to the application generally remains confined to areas where tight latency and throughput requirements exist, such as real-time control systems \cite{wanSurveyFPGABasedRobotic2021, podlubneModelBasedGenerationHardware2024}, and particle physics \cite{aggletonFPGABasedTrack2017a,sudvargFPGABasedDataProcessing2025}.

A key factor in the slow adoption of application-specific hardware designs is the so-called ``two-language problem,'' where scientific algorithms are usually prototyped and developed in high-level languages, but then must be translated into lower-level languages (such as RTL for FPGA designs) \cite{Julia}.
This leads to extra development effort and work, since two implementations generally need to be developed and maintained concurrently, and different teams might be responsible for the two layers.

Solving the two-language problem would mean taking the high-level implementation, written by the domain scientist/engineer, and directly compiling it into a hardware accelerator using High-Level Synthesis (HLS) tools.
Traditionally, most existing HLS tools have been based on a C/C++ ``high-level'' language and are based on compiler stacks developed for compiling and optimizing a stream of sequential operations, such as the LLVM framework.
While there has been success deploying HLS in fields such as video processing, graph processing and genomics, this traditional framework is not ideal for FPGA designs, and can lead to HLS-generated designs performing suboptimally and also non-portable source code \cite{congFPGAHLSToday2022}.

A key reason for this is a lack of high-level data about the design available to the HLS compiler when using the compiler stacks designed for sequential C/C++.
The intermediate representation (IR) used by these HLS compilers is at the operator level, with primitives such as addition, multiplication, bit shifts, etc.
This low-level IR loses information about higher-level parallel constructs and can prevent the compiler from identifying parallelism at the task/module level (e.g., sharing high-level operators like convolutions) \cite{YeHao:22}.
The designer can provide hints to the compiler with this missing information by adding pragmas to the code or using special libraries that encode operators, however these pragmas and libraries are generally toolchain-specific and non-portable.

Two approaches to overcome this lack of information have been proposed: starting from a higher-level language than C/C++ and raising the level of abstraction the IR in the compiler toolchain uses.
Compiling higher-level languages has been an active field of research for many decades, with prior works proposing toolchains for Haskell \cite{baaijClaSHStructuralDescriptions2010,gerardsHigherOrderAbstractionHardware2011}, Python \cite{haglundHardwareDesignScripting2003, cieszewskiPythonBasedHighlevel2014, jurkansPythonSubsetDigital2023, agostiniBridgingPythonSilicon2022, laiHeteroCLMultiParadigmProgramming2019}, Rust \cite{basuRustHardwareDescription2024} and Ruby \cite{lovicHDLRubyRubyExtension2023}.
These works varied in the integration with their language ecosystem, with some implementing custom domain-specific languages on top of the base language, such as HeteroCL \cite{laiHeteroCLMultiParadigmProgramming2019} and PyHDL \cite{haglundHardwareDesignScripting2003} for Python or HDLRuby \cite{lovicHDLRubyRubyExtension2023} for Ruby.
Other toolchains focus on reduced subsets of the language/framework, such as HPIPE \cite{hallTensorFlowGraphsLUTs2020} for compiling TensorFlow graphs to FPGA or MATLAB HDL coder.

Other HLS toolchains have proposed changing the level of abstraction for the IR used to generate the HDL.
The highest level works parsed the language directly and generated the HDL from the internal code and dataflow graph (CDFG) structures, such as \cite{cieszewskiPythonBasedHighlevel2014} for Python and \cite{BigMcI:22} for Julia.
More recently, toolchains have used existing higher-level IRs, such as SPIR-V \cite{jurkansPythonSubsetDigital2023} and Multi-level Intermediate Representation (MLIR) \cite{agostiniBridgingPythonSilicon2022,urbachHLSPyTorchSystem2022,YeHao:22}.

MLIR has emerged as the popular choice for building HLS tools currently, with the ability to generate HDL from MLIR provided by either writing SystemVerilog using the CIRCT framework \cite{CIRCT,urbachHLSPyTorchSystem2022}, generating C/C++ code for existing HLS toolchains \cite{YeHao:22,basalamaStreamHLSAutomaticDataflow2025a}, or emitting LLVM IR to feed into vendor-specific toolchains like Vitis HLS \cite{rodriguez-canalFortranHighLevelSynthesis2023a, liangSupportMLIRHLS2023}.
Prior works have generated MLIR using various front-ends, including walking the Clang syntax tree \cite{MosChe:21,YeJu:24}, the flang Fortran front-end \cite{rodriguez-canalFortranHighLevelSynthesis2023a}, and from PyTorch using PyTorch-MLIR \cite{YeJu:24,urbachHLSPyTorchSystem2022}.
The design of MLIR also allows for toolchains to implement their own IR ``dialects'' on top of the standard MLIR dialects to represent custom operations/features or to aid in optimisations.
Examples of custom dialects include Stencil-HMLS \cite{rodriguez-canalStencilHMLSMultilayeredApproach2023} for representing stencil computations, or HIR \cite{majumderHIRMLIRbasedIntermediate2024} to add scheduling information.

\subsection{Our Contribution}

In this work, we present an open-source and permissively licensed toolchain that compiles the Julia language, a language built for science and mathematics that includes native support for common math operations/concepts (e.g., linear algebra), into SystemVerilog by going through MLIR using the CIRCT framework.\footnote{Weblink to the repository has been removed for double-blind review}
Our toolchain --- called \textit{JuliaHLS} --- combines both the approaches mentioned above --- going from the Julia language straight to MLIR.
This allows our toolchain to have advanced analyses and optimisations, such as operator fusion for linear algebra or tiling, that were not previously possible without fragile `lifting passes' or programmer-annotated compiler directives.
We also leverage unique features of the Julia compiler infrastructure to build this tool, including the \texttt{AbstractInterpreter} interface and method table overlays.
The \texttt{AbstractInterpreter} interface allows for us to create a new \texttt{MLIRInterpreter} compiler flow that operates alongside the standard Julia compiler while reusing many parts from it (e.g., AST lowering, type analysis, optimisation passes, etc.) without relying on internals of the Julia compiler stack.

JuliaHLS is able to generate high-performance designs that correctly pass timing analysis at 100MHz and achieve up to 82.6\% the throughput of Dynamatic v2.0.
Additionally, in our testing, our toolchain was able to correctly compile several programs that Dynamatic was unable to, and was able to produce designs capable of running at higher clock frequencies. 

We begin by presenting necessary background information about Julia and relevant HLS toolchains in Section~\ref{sec:background}.
We then discuss the rationale for the design of JuliaHLS in Section~\ref{sec:design}, and its implementation in Section~\ref{sec:compiler}.
Several case studies are presented in Section~\ref{sec:results}, followed by a discussion of the results and future work in Sections~\ref{sec:conclusion} and~\ref{sec:future}. 


\begin{table*}[htbp!]
	\centering
	\fontsize{9pt}{11pt}\selectfont
    \setlength\extrarowheight{2pt}
    \begin{threeparttable}
	\begin{tabular}{>{\centering\arraybackslash} m{0.13\linewidth}
            >{\centering\arraybackslash} m{0.06\linewidth}
            >{\centering\arraybackslash} m{0.07\linewidth}
            >{\centering\arraybackslash} m{0.06\linewidth}
            >{\centering\arraybackslash} m{0.06\linewidth}
            >{\centering\arraybackslash} m{0.07\linewidth}
            >{\centering\arraybackslash} m{0.07\linewidth}
            >{\centering\arraybackslash} m{0.08\linewidth}
            >{\centering\arraybackslash} m{0.09\linewidth}
            >{\centering\arraybackslash} m{0.10\linewidth}}
        \toprule
		\multirow{2}{*}{\textbf{HLS Tool}} &
        \multirow{2}{\linewidth}{\textbf{Open source}} &
        \multirow{2}{\linewidth}{\textbf{Vendor\newline Agnostic}} &
        \multicolumn{2}{c}{\textbf{Scheduling}} &
        \multirow{2}{\linewidth}{\textbf{Usable\newline Designs}} &
        \multirow{2}{\linewidth}{\textbf{Reusable}} &
        \multicolumn{3}{c}{\textbf{Language}}\\
        \cline{4-5}\cline{8-10}
        & & & \textbf{Dynamic} & \textbf{Static} & & & \textbf{Source} & \textbf{IR} & \textbf{Target}\\\midrule
		\textbf{This work} & \cmark & \cmark & \cmark & \cmark & \cmark & \cmark & Julia & MLIR & SystemVerilog\\ \hline
        \citet{BigMcI:22} & \cmark & \cmark & \cmark & \xmark & \cmark/\xmark\tnote{3} & \xmark & Julia & Julia/\newline Dot graph & VHDL\\\hline
        \citet{LouGer:25} & \xmark & \xmark & \xmark & \cmark & \cmark & \cmark & Julia & MLIR & HLS C++\\\hline
        Matlab HDL Coder & \xmark & \cmark & \xmark & \cmark & \xmark & \xmark & Matlab & Unknown & Verilog/VHDL/\newline SystemVerilog\\\hline
		AMD Vitis HLS & \xmark & \xmark & \xmark & \cmark & \cmark &  \cmark/\xmark\tnote{2} & C/C++ & LLVM & Bitstream\\\hline
        Dynamatic \cite{JosGue:20}& \cmark & \cmark & \cmark & \xmark & \cmark \tnote{4}& \cmark/\xmark\tnote{2} & C/C++ & v1: LLVM\newline v2: MLIR\tnote{5} & VHDL\\\hline
        ScaleHLS \cite{YeHao:22} & \cmark & \xmark & \xmark & \cmark & \cmark & \cmark & C/C++\tnote{5}\newline PyTorch & MLIR & HLS C++ \\\hline
        HIDA \cite{YeJu:24} & \cmark & \xmark & \cmark & \xmark & \cmark & \cmark/\xmark\tnote{2} & C/C++\tnote{5}\newline PyTorch & MLIR & HLS C++ \\\hline
        PandA\newline Bambu HLS \cite{FerVit:21}& \cmark & \cmark & \xmark & \cmark & \cmark & \xmark & C/C++ & ``Bambu IR'' & Verilog/VHDL\\\hline
        LegUp HLS \cite{canisLegUpOpensourceHighlevel2013} & \cmark/\xmark\tnote{1} & \cmark/\xmark & \xmark & \cmark & \cmark & \cmark & C/C++ & LLVM & Verilog\\\hline
        \bottomrule
	\end{tabular}
    \begin{tablenotes}
    \item[1]{Initially open source, but now closed source},
	\item[2]{Partially reusable},
	\item[3]{Supports a very limited language subset},
	\item[4]{Toolchain makes additional assumptions},
    \item[5]{C/C++ is parsed to LLVM IR using Clang, and then raised to MLIR using Polygeist \cite{MosChe:21}}
    \end{tablenotes}
	\caption{Comparison of HLS toolchains and their features}
    	\label{tab:existing_work:comparison}
    \end{threeparttable}
    \vspace{-2\baselineskip}
\end{table*}




\section{Background}
\label{sec:background}

\subsection{The Julia Language}
\label{sec:background:julia}

Julia \cite{Julia} is an open-source, dynamically typed language built on top of the LLVM framework \cite{LatAdv:04} that allows writing code in a high-level, expressive and composable language and then using Just in time (JIT) compilation to create high-performance software.
The syntax closely resembles mathematical notation, and is similar to the MATLAB language, making it popular within the scientific computing and engineering communities.

Using Julia as the source language for HLS tools was first advocated for by \citet{BigMcI:22}, who presented several advantageous features, including:
\begin{description}

\item[High-level information]
Julia integrates mathematical concepts such as linear algebra as first-class features inside the language.
This allows the compiler to know about higher-level operations, as well as higher-level objects such as matrices, giving more opportunities for optimisation, parallelism and resource sharing.
This eliminates the need for lifting passes to reconstruct high-level information like what is done in tools like Polygeist \cite{JosGue:20,MosChe:21}.


\item[Compiler Infrastructure]
Julia's extensible compiler infrastructure (e.g., \texttt{AbstractInterpreter}) allows new target back-ends to integrate directly into standard distributions of the language and thus reuse existing infrastructure.
There have also been several other successful projects that target accelerator platforms from Julia, including GPUs \cite{BesVer:16, BesFok:19}, TPUs \cite{FisSab:18} and Graphcore IPUs \cite{Gio:23}.


\item[Ecosystem]
The Julia package ecosystem consists of many `pure Julia packages', which instead of simply providing a thin wrapper on top of a Fortran or C library, actually implement the algorithms in Julia.
This makes it possible to simulate custom numerics formats inside existing algorithms, and could also allow for accelerator cores to be created by just using the algorithms in existing packages instead of re-implementing them.


\item[Multiple dispatch and Type System]
Julia uses a dynamic type system that also allows for providing type constraints on variables and method arguments, and performs function calls using multiple dispatch, where the compiler examines all the method argument's types and selects the method that is most appropriate for the methond arguments.
This encourages the writing of generic code using abstract types and generic operators that is then specialized at compile time through a type-inference process to generate implementations tailored to specific data types.
The use of abstract types allows methods to be used with custom data types, such as arbitrary floating point or fixed point number formats (using the \texttt{ArbFloat} and \texttt{FixedPointNumbers} packages, respectively), allowing the functional simulation of algorithms to be implemented on FPGAs using CPU code instead of in a simulator/emulator.

\end{description}

\subsection{HLS Toolchains}

Many commercial and research HLS toolchains have been developed, with a high-level comparison of relevant toolchains in Table~\ref{tab:existing_work:comparison} and brief summaries below.

\begin{description}

\item[AMD Vitis]
is a commercial HLS tool that generates optimized hardware exclusively for AMD/Xilinx FPGAs.
It's design entry language is C/C++, however recent versions also support passing LLVM 7 IR directly.
Many research toolchains utilize Vitis as a back-end by either generating C/C++ (e.g., HIDA \cite{YeJu:24} and ScaleHLS \cite{YeHao:22}), or by passing LLVM IR directly (e.g., Fortran \cite{rodriguez-canalFortranHighLevelSynthesis2023a}).

\item[MATLAB HDL Coder]
is a commercial tool by MathWorks that compiles MATLAB code and Simulink models into IP cores using either Verilog, SystemVerilog or VHDL.
It can target many FPGA technologies, including AMD, altera and Microchip, however previous studies have shown that it produces designs with poor performance characteristics \cite{CurFio:23}.

\item[PandA Bambu HLS]
can compile C/C++ into statically scheduled, vendor-agnostic hardware designs \cite{FerVit:21}.
It has been extensively used by groups like the European Space Agency \cite{FerFio:24,LatFer:16,CecPal:12} to implement custom accelerators.

\item[ScaleHLS and HIDA]
are based on MLIR and can accept either C/C++ (parsed into LLVM IR using Clang, then raised to MLIR using Polygeist \cite{MosChe:21}) or PyTorch-MLIR designs.
HIDA \cite{YeJu:24} creates optimized dynamically scheduled designs, outputting C++ code to pass to vendor toolchains (e.g., Vitis).
ScaleHLS \cite{YeHao:22} creates statically scheduled designs, and outputs either C++ or LLVM IR to pass to further toolchains.

\item[Dynamatic]
produces dynamically scheduled VHDL designs from C/C++ inputs.
The initial Dynamatic tool \cite{JosGue:20} used LLVM IR for internal optimisation passes, however Dynamatic v2.0 has been rewritten to use MLIR for internal optimisation passes (parsing C/C++ via Polygeist \cite{MosChe:21}).

\item[Julia HLS Toolchains]
- Initial work by \citet{BigMcI:22} modified the Julia compiler to extract the CDFG, and then converted it into a dot graph of elastic components to pass to Dynamatic for RTL generation.
It only supported a very small subset of the Julia language (and did not support matrices/linear algebra), and was unmaintainable due to its close coupling to the Julia compiler internals.

\citet{LouGer:25} recently proposed an MLIR-based toolchain to compile Julia into C++ for Vitis HLS, converting the Julia typed IR directly into MLIR Structured Control Flow (\textit{scf}) dialect for processing by ScaleHLS.
This workflow converts matrices directly into low-level memory buffer (\texttt{memref}) statements, losing higher-level information about the type of matrix and subsequent operations.


\end{description}

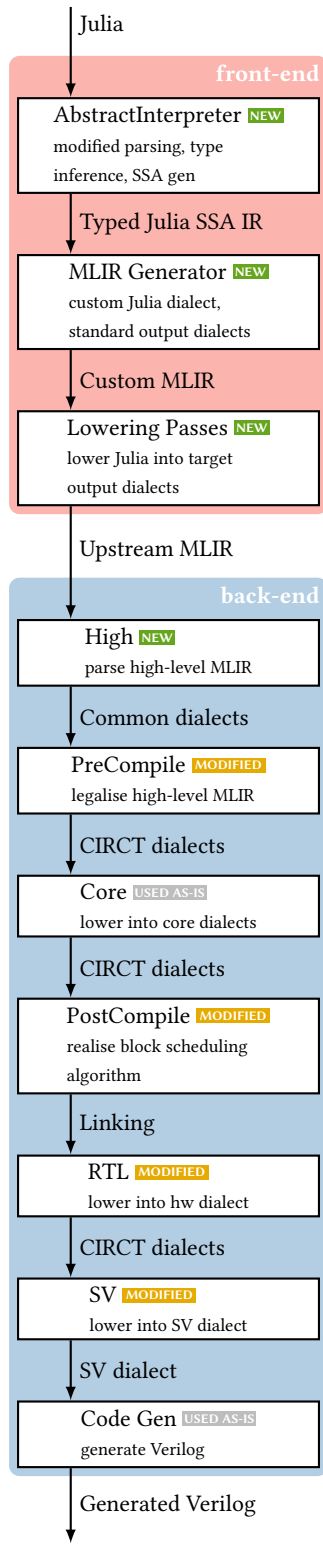
\begin{figure}
\begin{tikzpicture}

\node[pass] (n1) {AbstractInterpreter \NEW{} \\ \footnotesize modified parsing, type \\ \footnotesize inference, SSA gen};

\node[pass, below=\defaultgap of n1] (n2) {MLIR Generator \NEW{} \\ \footnotesize custom Julia dialect, \\ \footnotesize standard output dialects};

\node[pass, below=\defaultgap of n2] (n3) {Lowering Passes \NEW{} \\ \footnotesize lower Julia into target \\ \footnotesize output dialects};

\node[pass, below=15mm of n3] (n4) {High \NEW{} \\ \footnotesize parse high-level MLIR};

\node[pass, below=\defaultgap of n4] (n5) {PreCompile \MODIFIED{} \\ \footnotesize legalise high-level MLIR};

\node[pass, below=\defaultgap of n5] (n6) {Core \ASIS{} \\ \footnotesize lower into core dialects};

\node[pass, below=\defaultgap of n6] (n7) {PostCompile \MODIFIED{} \\ \footnotesize realise block scheduling \\ \footnotesize algorithm};

\node[pass, below=\defaultgap of n7] (n8) {RTL \MODIFIED{}  \\ \footnotesize lower into hw dialect};

\node[pass, below=\defaultgap of n8] (n9) {SV \MODIFIED{} \\ \footnotesize lower into SV dialect};

\node[pass, below=\defaultgap of n9] (n10) {Code Gen \ASIS{} \\ \footnotesize generate Verilog};

\draw[arr] ([xshift=\arrshift, yshift=12mm]n1.north) to node[auto, pos=0.2] {Julia} ([xshift=\arrshift]n1.north);
\draw[arr] ([xshift=\arrshift]n1.south) to node[auto] {Typed Julia SSA IR} ([xshift=\arrshift]n2.north);
\draw[arr] ([xshift=\arrshift]n2.south) to node[auto] {Custom MLIR} ([xshift=\arrshift]n3.north);
\draw[arr] ([xshift=\arrshift]n3.south) to node[auto, pos=0.4] {Upstream MLIR} ([xshift=\arrshift]n4.north);
\draw[arr] ([xshift=\arrshift]n4.south) to node[auto] {Common dialects} ([xshift=\arrshift]n5.north);
\draw[arr] ([xshift=\arrshift]n5.south) to node[auto] {CIRCT dialects} ([xshift=\arrshift]n6.north);
\draw[arr] ([xshift=\arrshift]n6.south) to node[auto] {CIRCT dialects} ([xshift=\arrshift]n7.north);
\draw[arr] ([xshift=\arrshift]n7.south) to node[auto] {Linking} ([xshift=\arrshift]n8.north);
\draw[arr] ([xshift=\arrshift]n8.south) to node[auto] {CIRCT dialects} ([xshift=\arrshift]n9.north);
\draw[arr] ([xshift=\arrshift]n9.south) to node[auto] {SV dialect} ([xshift=\arrshift]n10.north);
\draw[arr] ([xshift=\arrshift]n10.south) to node[auto] {Generated Verilog} ([yshift=-1cm,xshift=\arrshift]n10.south);

\node[above=2mm of n1.north east, anchor=south east, text=colorFRONTENDfg, inner sep=0pt] (frontend_lbl) {\bf front-end};

\node[above=2mm of n4.north east, anchor=south east, text=colorBACKENDfg, inner sep=0pt] (backend_lbl) {\bf back-end};

\begin{scope}[on background layer]
\node[fit=(n1)(n2)(n3)(frontend_lbl), frontend] (frontend) {};
\node[fit=(n4)(n5)(n6)(n7)(n8)(n9)(n10)(backend_lbl), backend] (backend) {};
\end{scope}

\end{tikzpicture}

\caption{High-level system architecture, with each pass tagged as \NEW{} for this project, \MODIFIED{} from CIRCT, or simply \ASIS{} from CIRCT.}
\label{fig:full_system}
\Description[High Level System Architecture]{The high level flow goes from Julia into the front end. This lowers it into upstream MLIR, passes it into the back end and generates Verilog.}
\end{figure}

\section{Toolchain Design}
\label{sec:design}

The Julia language was chosen as the source language for this HLS toolchain because of its focus on science and engineering computations, and the four points outlined in Section~\ref{sec:background:julia}.
A key vision for JuliaHLS is that compiling an FPGA implementation of a function should not require any changes to the algorithm's implementation, provided the algorithm is implemented in a generic form using abstract types.
To enable this, we do not implement a domain-specific language to describe the accelerator in the Julia language, but instead directly compile the pure Julia code into the RTL for the hardware accelerator.

\subsection{Intermediate Representation}

The first HLS project in Julia by \citet{BigMcI:22} attempted to keep the design in the Julia typed SSA IR as long as possible, only handing off the design to Dynamatic v1 as a dot graph for the HDL generation.
This led to problems with performance, though, since the toolchain could not leverage any pre-existing HLS optimisation passes, with the evaluation showing a large increase in resource usage when using the Julia toolchain instead of Dynamatic v1.
To overcome this, the authors proposed rewriting the optimisation passes from the LLVM-based toolchains into Julia to run on the Julia typed SSA IR before generating the HDL, and then also generating the HDL directly from Julia to create a completely Julia-based toolchain.
This suggestion of building a completely Julia-based toolchain with all the components and optimisation passes written in Julia is unmaintainable though and would isolate the JuliaHLS toolchain development from the wider HLS community.

Other Julia toolchains for offloading to accelerators like GPUs \cite{BesVer:16, BesFok:19} and the Graphcore IPU \cite{Gio:23} leverage the fact Julia is LLVM-based to intercept the LLVM IR and redirect it to the LLVM target back-end for the desired accelerator.
While a similar flow could be used for RTL generation by taking the Julia-generated LLVM IR and passing it into AMD Vitis HLS (through a version down-converter, similar to what was done for Fortran HLS \cite{rodriguez-canalFortranHighLevelSynthesis2023a}), this flow would be very vendor-specific and the down-converting of LLVM into a version Vitis can use could become more difficult over time.

Quite a lot of the information about higher-level constructs, like matrices and their operations, that is present in the Julia typed SSA IR gets removed by the Julia to LLVM IR lowering pass, which would hide that information from the HLS toolchain.
Additionally, LLVM IR models the program in an imperative and procedural manner, which is insufficient for capturing the inherently parallel nature of hardware, reducing the compiler's ability to generate efficient hardware designs.
This is termed ``premature lowering'', and Lattner et al. go as far as saying that ``Premature lowering is the root of all evil'' \cite{LatAmi:21}, emphasising that LLVM is insufficient for HLS.

Instead, JuliaHLS uses the MLIR ecosystem as the core of the toolchain, and uses MLIR dialects as its intermediate representation.
MLIR is built as a modular framework consisting of a set of core `dialects' that define sets of operations, attributes and types in a single namespace.
Unlike LLVM, which only allows a single dialect and can't natively represent nested code or generic loops, MLIR allows for dialects to be overlaid onto each other in the same generated IR, allowing the mixing of higher-level operations (e.g., matrix addition) with lower-level operations (e.g., scalar addition), and also to preserve program structure and control flow.
MLIR contains many different dialects, including \textit{scf} for structured control flow contructs like \texttt{for} and \texttt{if}, \textit{linalg} for representing matrix/vector operations, and \textit{tosa} for representing generic tensors.


\subsection{HDL Generation}

There are two main routes that an MLIR-based HLS toolchain can use to lower a program into HDL: lower the MLIR to the \texttt{llvm} dialect or transpile to C via the \texttt{emitc} dialect and use a tool like AMD Vitis HLS to generate the RTL, or to use an MLIR-based HLS tool with HDL output, like CIRCT \cite{CIRCT} or Dynamatic v2.0 \cite{JosGue:20}.
The main problem with the first option is that it relies on transferring data to another tool, like Vitis HLS, using a representation (either LLVM IR or C/C++ code) at a different level.
This flow can also introduce vendor lock-in, since these external tools that take LLVM/C/C++ code usually requires specific code structure or libraries to identify constructs or interfaces, and many such tools rely heavily on pragmas and other directives to control the design-flow and generate efficient designs and interfaces, which can be tool dependent.


JuliaHLS uses the CIRCT \cite{CIRCT} infrastructure to keep the design in MLIR during the entire process, and then lower to the SystemVerilog dialect before writing out the SystemVerilog description of the design.
CIRCT is a collection of MLIR dialects, conversion and optimisation passes that are focused on hardware design, initially starting as a dataflow accelerator design tool, but also recently integrating initial support for statically scheduled designs using the Calyx toolchain \cite{NigTho:21}.



\section{The Compiler Stack}
\label{sec:compiler}

JuliaHLS implements a 3-stage modular compiler architecture, shown in Figure~\ref{fig:full_system}, where the front-end extracts the program, the first half of the back-end optimises and transforms the program, and the latter half generates the SystemVerilog.

\subsection{Front-end}

The main role of the JuliaHLS front-end is to extract a type-stable CDFG from the idiomatic Julia source code and convert it into MLIR.
This conversion is broken down into three stages: resolving type-stability, resolving mismatches between the IRs and generating optimised MLIR consisting of standard dialects.
A high-level overview of the architecture of this component is shown in the upper half of Figure~\ref{fig:full_system}.


\subsubsection{Type Stability}
\label{sec:design:front_end:type_stability}

Since MLIR is statically and strongly typed while idiomatic Julia does not have to be, the first compilation phase is to extract a statically and strongly typed version of the Julia code, known as a ``type stable'' code.
There are two key steps required for generating type stable code for the JuliaHLS toolchain to use: perform analysis to annotate operators and variables with a single concrete type, and eliminating runtime polymorphism (i.e., dynamic dispatch) from the code.
Eliminating runtime polymorphism and only having compile-time polymorphism (i.e., static dispatch) is essential for generating efficient hardware, as dynamic dispatch can result in exponentially increasing design-sizes due to additional hardware and control flow, leading to unpredictable characteristics or wasted resources.
This, however, poses an inherent challenge because the standard Julia compiler type-inference phase only produces a best-effort attempt at type-stability and quits early when it determines the performance uplift of knowing the types statically on a fetch-execute processor is deemed to be insignificant. 
Other Julia toolchains, like the GPU compiler \cite{BesVer:16}, do not strictly enforce strong type stability or static dispatch, and instead will leverage the close coupling between CPUs and GPUs to delegate type-unstable sections of code to the CPU and type-stable parts to the GPU.

In JuliaHLS, we enforce a requirement for only static dispatch and type stable code, similar to the requirements of the recent Julia binary compilation workflow \cite{CarTap:25}.
We do this by creating a new \texttt{MLIRInterpreter} compiler flow that is a concrete implementation of Julia's `AbstractInterpreter', which allows JuliaHLS to modify the behavior of the internal Julia compiler flow without having to patch internal code.
To enforce static dispatch, Julia's `method table overlays' feature is used to inject special versions of problematic methods in order to provide concrete implementations into just the \texttt{MLIRInterpreter} compiler, leaving the methods used by the standard CPU or GPU back-ends untouched.
Additionally, \texttt{MLIRInterpreter} introduces custom inlining policies that allow us to extract high-level or generic program intrinsics when appropriate.
For example, Julia's default CPU-optimised inlining policy will assume sequential execution of linear algebra operations and may inline certain computations on vectors/matrices.
However, to allow for more optimisations to be performed in the HLS toolchain, the matrix operations should be extracted directly as a single operation that can then be passed into the back-end.

The end result is a type-stable high-level representation of the original program in the Julia typed SSA IR, which uses basic blocks to model the control-dependencies and SSA for data-dependencies.

\subsubsection{Generating MLIR}

The next step is to take the Julia IR produced by the previous stage and to translate it to very high-level MLIR.
This is split into a two-step process that first directly translates the Julia IR into a custom MLIR dialect built for JuliaHLS, known as the \texttt{julia} dialect, before then applying transformation passes to resolve IR-level mismatches and create idiomatic MLIR.
Separating the conversion into two stages allows us to prioritize a simple and correct code generation infrastructure because this reduces the overall complexity of the conversion routines, and allows for better compartmentalization of the required clean-ups because they can be implemented as independent MLIR transformation passes.


The MLIR passes that implement these program transformations are written in Julia and then registered within the MLIR engine as an external pass.
The main transformations applied are:
\begin{itemize}
    \item converting Julia's column-wise arrays to row-wise tensor objects,
    \item mapping Julia SSA violations to SSA-legal equivalents (e.g. converting Julia's memory-buffer model to MLIR's high-level tensor object),
    \item converting Julia's 1-based array indexing to 0-based,
    \item additional type-casting, and
    \item mapping math operations found in the Julia standard library to FPGA-legal alternatives.
\end{itemize} 

The next step is to lower the \texttt{julia} dialect into standard MLIR dialects commonly supported by downstream back ends like LLVM-based JITs or the CIRCT framework.
This is achieved by writing custom passes in Julia that efficiently lower the \texttt{julia} dialect into the \texttt{TOSA}, \texttt{tensor}, \texttt{arith} and \texttt{cf} MLIR dialects.
The use of the \texttt{TOSA} and \texttt{tensor} dialects allows for JuliaHLS to directly model linear algebra objects and operations within the generated IR, in contrast to prior works such as \cite{LouGer:25}, which model matrices/arrays as a shared buffer using the \texttt{memref} dialect that individually controls the access of elements, which is analogous to buffers in both software and hardware.
These matrix operations were preserved in the Julia typed SSA IR using the \texttt{MLIRInterpreter} custom inlining logic, allowing us to extract expressions and linear algebra from the program directly, and thus make use of the sophisticated and well-maintained MLIR dialects and passes to optimise the program before lowering it to a bufferised representation.
As a result, our toolchain allows the hardware back-end to apply domain-specific heuristics for hardware-friendly implementations.

\subsubsection{Optimisations}

The standard Julia compiler generates very verbose typed SSA IR as it offloads the majority of its optimisations to its LLVM back-end.
This means that the MLIR generated from \texttt{MLIRInterpreter} can be verbose and unoptimized, so the JuliaHLS MLIR back-end must also apply the same optimisation passes --- which are already implemented in the standard MLIR library.

The main difference between the Julia LLVM optimisation passes and the MLIR optimisation passes used by JuliaHLS is that the default control-flow simplification applied by the `canonicalize' pass must be reduced from `aggressive' to `normal' for JuliaHLS, since aggressive could lead to control-flow being folded so aggressively that CIRCT cannot determine whether the CDFG would result in a deadlock when generating a dataflow circuit.
For example, if the execution pipeline splits into many different concurrent streams sharing different hardware units, then this either requires an excessively complex control-flow network to enact sufficient back-pressure (very high overheads and error-prone), or the tool must provide a guarantee that the resources can be safely shared.
As a result, the framework provides a blanket ban on all `irregular' control flow to mitigate this problem.
Additional passes are also written and applied to `reformat' the MLIR into the subset supported by CIRCT, i.e. no static variables. 
In practice, this component was crucial for synthesising large programs to ensure that they both fit on an FPGA and have performance characteristics comparable to state-of-the-art toolchains, as we will show in Section~\ref{sec:results}.


\subsection{Back-end}

The JuliaHLS back-end component takes the MLIR generated by the front-end and uses the CIRCT library to generate RTL designs specified using Verilog or SystemVerilog.
We combine a selection of existing (unmodified) dialects and transformations from CIRCT with several custom passes and some direct modifications to the CIRCT library to generate RTL designs that are able to pass timing at 100MHz and are compatible with a range of vendor synthesis tools, including those from altera and AMD.

This back-end contains six layers of abstraction:
\begin{enumerate}
    \item \textit{High}: high-level MLIR dialects and operations,
    \item \textit{PreCompile}: lowers to MLIR dialects supported by CIRCT,
    \item \textit{Core}: transforms the program to exclusively use operations legal in CIRCT,
    \item \textit{PostCompile}: performs either static or dynamic scheduling using \texttt{calyx} \cite{NigTho:21} and \texttt{handshake} \cite{PetPao:22} respectively,
    \item \textit{RTL}: lowers an RTL circuit, and
    \item \textit{sv}: generates SystemVerilog in MLIR format
\end{enumerate}
Each abstraction layer is connected by a pipeline of passes that progressively lowers the design into RTL, and finally HDL.

The \textit{High-to-PreCompile} pipeline is unique to JuliaHLS and is built to handle lowering the linear algebra operations, while the \textit{PreCompile-to-Core}, \textit{Core-to-PostCompile} and \textit{PostCompile-to-RTL} pipelines are based on modified upstream CIRCT passes to ensure that the generated RTL was correct and synthesisable (e.g., fit on small FPGAs and run at 100MHz).


\subsubsection{High-to-PreCompile pipeline}

This pipeline is unique to JuliaHLS and will lower the high-level MLIR dialects into the dialects accepted by CIRCT, specifically to \texttt{arith}, \texttt{memref}, \texttt{func}, \texttt{affine}, and either \texttt{cf} for dataflow generation or \texttt{scf} for static scheduling using Calyx.
The most important component of this pipeline is the \textit{TosaToLinalg} transformation, which first lowers \texttt{tosa} to \texttt{linalg} and then bufferises the program (i.e., lowers SSA-compliant tensor objects into non-SSA memory buffers).
Of note is that one limitation of globally bufferising a program along the function boundaries is that it is no longer valid when recursion is used, however, JuliaHLS does not support recursion, as it does not map to efficient and reliable hardware. 
 
The next step in \textit{TosaToLinalg} is to apply a new pass, known as \texttt{OutputMemrefPassByRef}, that will convert functions returning n-dimensional arrays to instead pass in a reference to the output array as an argument.
This modification is needed to ensure that CIRCT generates a Load-Store Queue (LSQ) and allows ingress/egress of data to memory outside the design, i.e., off-chip DDR3.

The final step is to apply a modified version of the \texttt{LowerLinalg} \texttt{ToAffine} pass and then the \texttt{LowerAffinePass} to create a control-flow representation of the matrix operations.
The pass requires modification in JuliaHLS because the original pass greedily applies optimisation patterns for all registered dialects and thus excessively folds control-flow.
However, CIRCT dataflow generation does not support excessively folded control flow, because this implies that resources can be shared between branches, and the ‘handshake’ dialect cannot yet provide a formal guarantee that this does not result in a deadlock.

\subsubsection{PreCompile-to-Core pipeline}

This lowering pipeline will transform the MLIR from the \textit{PreCompile} stage into MLIR accepted by either the dataflow generator in CIRCT or the static scheduler in Calyx.
For static scheduling, this pipeline is a single pass from CIRCT called \texttt{SCFToCalyxPass}, which converts the structured control flow into the \texttt{calyx} dialect for static scheduling.
For dynamic scheduling, this pipeline has seven passes, with the last five passes from the standard CIRCT \texttt{handshake} dialect lowering pipeline, but with a new pass called \texttt{DeleteUnusedMemory}, and a modified \texttt{InsertMergeBlocksPass} at the beginning of the pipeline.

The \texttt{DeleteUnusedMemory} pass analyses the program to find unnecessary data movement in the design by examining the read- and write-effects of memory buffers, and then delete unnecessary copy operations, buffers and memories.
This pass is key to ensuring that memory- and maths-heavy programs have both competitive performance characteristics and fit on budget FPGAs, such as the Pynq Z1, and we have seen an average performance improvement of 25-40\% when using this pass.

JuliaHLS also contains improvements to the \texttt{InsertMergeBlocks} pass in the standard CIRCT library, which is a pass to insert additional `merge' blocks in the CDFG to ensure that a single block has only two backedges.
Specifically, we have modified this pass to handle the case where a block only has one back-edge, and modified it to only insert the additional merge blocks when truly needed instead of just everywhere.
These changes have reduced the complexity of the control-data network, and on average reduced the resource usage of the synthesised RTL by up to 20\%.


\subsubsection{Core-to-PostCompile pipeline}

When using dynamic scheduling, this pipeline will materialize the fork and sink operations that connect the various nodes of the CDFG, and then insert the various buffers required in the design.
When using static scheduling, this pipeline will begin by lowering the control schedule computed by Calyx into a finite state machine (FSM), and then connect the various control ports (i.e., clock signals, start flags and reset wires) of the components to the top-level control ports.

\subsubsection{RTL-to-SV pipeline}

This pipeline transforms the scheduled RTL design into the \texttt{sv} dialect, which is used to actually emit the SystemVerilog for the design.

For dynamically scheduled designs, JuliaHLS contains a modified version of the \texttt{HWLoweringPipeline} pass, which maps operations to dynamically generated IP cores.
By default, this pass would instantiate LUT-based RAM with an asynchronous output, however FPGA synthesis tools only map memory to FPGA primitives (such as BRAM) if they have synchronous outputs (i.e., have a registered output).
LUT-based memories are often too slow to meet fast timing requirements (e.g., 100 MHz), so we have modified this pass to instantiate a register on the output path, which improves the synthesisability of the generated hardware at the cost of an additional access cycle for memory reads.
This also reduced the fan-out from heavily used LUTs, thus decreasing the overall load on the FPGA interconnect fabric.

For statically scheduled designs, this pipeline converts the FSM into SystemVerilog and then links it with the relevant IP core for the computation.
JuliaHLS contains minor modifications to the original CIRCT/Calyx passes to ensure they create synthesisable SystemVerilog code and respect the configuration in CIRCT's Verilog generation infrastructure.


\subsection{Memory}


JuliaHLS supports memory operations for dynamically scheduled designs.
The underlying \texttt{handshake} dialect uses a shared memory model with the ability to either generate on-chip RAM instances or to communicate with external memory via LSQs.
External memory support is implemented using a parameterisable LSQ to AXI4-Stream adapter that manages each read-write channel via a round-robin arbiter, and preserves data dependencies by exerting back-pressure into the dataflow circuit to pause execution until the request has been successfully completed.
This AXI core cleanly integrates with both simulators and on real hardware such as a Pynq Z1 FPGA using the AMD DataMover IP core to read and write to external DDR3 memory.

\subsection{Julia Integration}

The JuliaHLS back-end is written in C++ and contains two main design flows: \texttt{HLSToolDynamic} and \texttt{HLSToolStatic}, for dynamically scheduled and statically scheduled HDL, respectively.
Since Julia can interface with C better than C++, a thin C API wrapper was created around these two flows, that is then compiled into a shared library.
Julia accesses the back-end toolflows using bindings autogenerated by Clang.jl\footnote{https://github.com/JuliaInterop/Clang.jl} that expose the various options and some basic functions to create and run the toolflows.

The full compilation process is abstracted with a Julia package,
which can be directly installed from the Julia REPL.\footnote{This package has been redacted for double-blind review and will be included in the camera-ready version.}
It wraps the Julia-based front-end with the back-end to provide a black box compilation process that can be invoked and customised directly from the REPL or within Julia scripts.
The RTL generation process is also customisable via CIRCT's export options to allow the RTL to taget a wide range of synthesis tools, including altera Quartus, AMD Vivado and Verilator.

\section{Case Studies}
\label{sec:results}

To demonstrate the JuliaHLS toolflow, we present two case studies that highlight various aspects of the toolchain: an implementation of a CORDIC algorithm and a 2D convolution kernel.
Aside from these examples, we have also tested the toolchain on others, ranging from simple operations and control flow to an implementation of the Newton-Raphson method and a complicated ODE solver for boundary value problems.

\subsection{Case study descriptions}
\label{sec:results:case_studies}


\subsubsection{Cordic}

This case study applies ten stages of the CORDIC algorithm \cite{Vol:59} to iteratively evaluate the cosine function.
This example is implemented using fixed-point numbers, which are provided by a Julia package called \texttt{FixedPointNumbers}, and highlights the use of method table overlays to add support for new types to JuliaHLS.
This program also explicitly demonstrates support for a wide range of language features, including control flow, ‘Look-Up-Tables’ via array indexing, bit operations, integer arithmetic, and generic methods.
A snippet of this program is shown in Listing \ref{fig:coordic:loop}.


\begin{listing}[t!]
\begin{minted}[fontsize=\small,frame=single,highlightlines={},highlightcolor=orange!40]{julia}
function cordic(theta::Fixed{T,N}, K::Fixed{T,N}) where {T,N}
    # arctangent lookup table 
    angles = @MMatrix [
       Fixed{T,N}(atan(1));
       Fixed{T,N}(atan(0.5));
    ...
    ]
    
    x = K
    y = Fixed{T,N}(0)
    z = theta

    for i in one(T):T(length(angles))
        di::Fixed{T, N} = z ≥ Fixed{T, N}(0) ?
            Fixed{T, N}(1) : Fixed{T, N}(-1)
        shift::T = i - one(T)

        # cordic rotation updates
        x_new = x - (di * (y >> shift))
        y_new = y + (di * (x >> shift))
        z_new = z - di * angles[i]

        x, y, z = x_new, y_new, z_new    
    end

    return x
end
\end{minted}
\caption{Snippet of the CORDIC case study.}
\label{fig:coordic:loop}
\end{listing}

\subsubsection{Conv2d\_im2col}

This case study performs a two-dimensional convolution using the ‘im2col’ algorithm, with the program shown in Listing~\ref{fig:im2col}.
We instantiate this algorithm with a 3x3 input matrix and a 2x2 kernel, but it can be extended to arbitrarily large vector sizes.
The main features demonstrated by this example are fully parameterisable accelerator development, matrix operations and manipulations, preservation of Write After Read (WAR) and Read After Write (RAW) dependencies, returning matrices, instantiating large on-chip memories, interacting with external memory and highly complex control-flow.

\begin{listing}[t!]
\begin{minted}[fontsize=\small,frame=single,highlightlines={},highlightcolor=orange!40]{julia}
function conv2d_im2col(A::MMatrix{MA,NA,T},
         K::MMatrix{KH,KW,T}) where {MA,NA,KH,KW,T}
  OM = MA - KH + 1
  ON = NA - KW + 1

  # im2col dims
  R = KH * KW
  C = OM * ON

  # build the patch matrix
  S = MMatrix{R, C, T}(undef)
  for j in 1:ON, i in 1:OM, v in 1:KW, u in 1:KH
    row = (v-1)*KH + u
    col = (j-1)*OM + i
    @inbounds S[row, col] = A[i+u-1, j+v-1]
  end

  # build the flattened representation
  k = MMatrix{R, 1, T}(undef)
  for v in 1:KW, u in 1:KH
    idx = (v-1)*KH + u
    @inbounds k[idx, 1] = K[u, v]
  end

  ycol = S' * k   # GEMM

  # scatter back into Y
  Y = MMatrix{OM, ON, T}(undef)
  for j in 1:ON, i in 1:OM
    idx = (j-1)*OM + i
    @inbounds Y[i, j] = ycol[idx, 1]
  end

  return Y
end
\end{minted}
\caption{2D convolution kernel.}
\label{fig:im2col}
\end{listing}




\begin{listing}[t!]
\begin{minted}[fontsize=\small,frame=single,highlightlines={},highlightcolor=orange!40]{julia}
function implicit_else(A, B)
   result = 0
   if A < B
      result = A + B
   elseif A > B
      result = A - B
   end
   return result
end
\end{minted}
\caption{\texttt{implicit\_else} testcase.}
\label{fig:implicitelse}
\end{listing}

\subsection{Dynamic and Static Scheduling}

Unfortunately, due to the Calyx tool and dialect being only recently integrated with CIRCT, it is unable to support memories or synthesize loops.
Instead, to demonstrate the ability for JuliaHLS to create both dynamically and statically scheduled code, we use a simpler example of an if statement that contains an implicit else clause, given in Listing~\ref{fig:implicitelse}.

This program is compiled with both scheduling strategies to examine performance and resource usage, with the results shown in Table~\ref{tab:implicit_else}.
The RTL for both designs was then synthesised using AMD Vivado targeting a Pynq Z1 FPGA to perform timing analysis, with both successfully passing at a clockspeed of 100 MHz.
Note that the latency is very similar for the two designs, however the statically scheduled design achieves a nearly $8\times$ reduction in resource utilization.

\begin{table}[t!]
\centering
\begin{tabular}{|l|l|r|r|r|}
\hline
& \textbf{Latency (Cycles)} & \textbf{LUT/ALM} & \textbf{FF} & \textbf{DSP} \\
\hline
\textit{Dynamic} & 13 & 2.29\% & 2.52\% & 0.0\%\\
\textit{Static}  & 14 & 0.27\% & 0.14\% & 0.0\%\\
\hline
\end{tabular}
\caption{Synthesis results for the scheduling comparison.}
\label{tab:implicit_else}
\vspace{-2\baselineskip}
\end{table}


\subsection{Toolchain Performance}
\label{sec:results:scientific_programs}

In this section, we analyse how JuliaHLS compiles the two scientific case studies presented in Section~\ref{sec:results:case_studies} using the dynamic scheduling toolflow only, since the static scheduling toolflow using Calyx does not support many features used by these examples.

\subsubsection{Program complexity}

\begin{table*}[ht!]
	\centering
	\fontsize{9pt}{11pt}\selectfont
    \setlength\extrarowheight{2pt}
	\begin{tabular}{>{\centering\arraybackslash} m{0.05\linewidth}
            >{\centering\arraybackslash} m{0.09\linewidth}|
            >{\centering\arraybackslash} m{0.06\linewidth}
            >{\centering\arraybackslash} m{0.04\linewidth}
            >{\centering\arraybackslash} m{0.04\linewidth}|
            >{\centering\arraybackslash} m{0.06\linewidth}
            >{\centering\arraybackslash} m{0.04\linewidth}
            >{\centering\arraybackslash} m{0.04\linewidth}|
            >{\centering\arraybackslash} m{0.06\linewidth}
            >{\centering\arraybackslash} m{0.04\linewidth}
            >{\centering\arraybackslash} m{0.04\linewidth}|
            >{\centering\arraybackslash} m{0.06\linewidth}
            >{\centering\arraybackslash} m{0.04\linewidth}
            >{\centering\arraybackslash} m{0.04\linewidth}}
        \toprule
		\multirow{2}{*}{\textbf{Design}} &
        \multirow{2}{\linewidth}{\textbf{Component}} &
        \multicolumn{3}{c}{\textbf{Basic Blocks}} &
        \multicolumn{3}{c}{\textbf{Total Ops}} &
        \multicolumn{3}{c}{\textbf{Arithmetic Ops}} &
        \multicolumn{3}{c}{\textbf{Memory Copies}}\\
        \cline{3-5}\cline{6-8}\cline{9-11}\cline{12-14}
        & & \textbf{No opt} & \textbf{Opt} & 
        & \textbf{No opt} & \textbf{Opt} & 
        & \textbf{No opt} & \textbf{Opt} &
        & \textbf{No opt} & \textbf{Opt}\\\midrule
		\multirow{2}{*}{cordic}
        & Front-end & 14 & 7 & \textbf{2$\times$}   & 129 & 58 & \textbf{2.22$\times$} & 44 & 17 & \textbf{2.59$\times$} & 3 & 2 & \textbf{1.5$\times$}\\
        & Back-end  & 14 & 5 & \textbf{2.8$\times$} &  88 & 76 & \textbf{1.16$\times$} & 18 & 18 & -                     & 4 & 1 & \textbf{4$\times$}\\\midrule
        \multirow{2}{*}{conv2d}
        & Front-end & 49 & 49 & -                     & 486 & 379 & \textbf{1.28$\times$} & 70 & 56 & \textbf{1.23$\times$} & 84 & 84 & -\\
        & Back-end  & 79 & 63 & \textbf{1.25$\times$} & 306 & 203 & \textbf{1.51$\times$} & 82 & 78 & \textbf{1.05$\times$} & 87 &  6 & \textbf{14.5$\times$}\\
        \bottomrule
	\end{tabular}
	\caption{Program complexity with and without optimisations in the component of JuliaHLS.}
    \label{tab:scires_complexity}
    \vspace{-1\baselineskip}
\end{table*}

\begin{table*}[ht!]
\centering
\begin{subtable}[t]{0.45\textwidth}
\centering
\begin{tabular}{|l|c|c|c|}
\hline
& \textbf{Unoptimized} & \textbf{Optimized} & \textbf{Speedup}\\\hline
cordic & 5765 & 241 & \textbf{23.92$\times$}\\\hline
conv2d & >1,000,000,000 & 1946 & \textbf{$\infty$}\\\hline
\end{tabular}
\caption{Latency in Cycles (lower is better)}
\label{tab:performance_analysis:latency}
\end{subtable}
\hspace{0.05\textwidth}
\begin{subtable}[t]{0.45\textwidth}
\centering
\begin{tabular}{|l|c|c|c|}
\hline
& \textbf{Unoptimized} & \textbf{Optimized} & \textbf{Speedup}\\\hline
cordic & 17346.05 & 414937.76 & \textbf{23.92$\times$}\\\hline
conv2d & $\approx 0$ & 17346.05 & \textbf{$\infty$}\\\hline
\end{tabular}
\caption{Throughput in Operations per Second (higher is better)}
\label{tab:performance_analysis:throughput}
\end{subtable}
\caption{Performance characteristics}
\label{tab:performance_analysis}
\vspace{-1\baselineskip}
\end{table*}

\begin{table*}[ht!]
	\centering
	\fontsize{9pt}{11pt}\selectfont
    \setlength\extrarowheight{2pt}
	\begin{tabular}{>{\centering\arraybackslash} m{0.04\linewidth}
            >{\centering\arraybackslash} m{0.10\linewidth}|
            >{\centering\arraybackslash} m{0.08\linewidth}
            >{\centering\arraybackslash} m{0.08\linewidth}
            >{\centering\arraybackslash} m{0.05\linewidth}|
            >{\centering\arraybackslash} m{0.08\linewidth}
            >{\centering\arraybackslash} m{0.08\linewidth}
            >{\centering\arraybackslash} m{0.05\linewidth}|
            >{\centering\arraybackslash} m{0.08\linewidth}
            >{\centering\arraybackslash} m{0.08\linewidth}
            >{\centering\arraybackslash} m{0.05\linewidth}}
        \toprule
		\multirow{2}{*}{\textbf{Design}} &
        \multirow{2}{*}{\textbf{Toolchain}} &
        \multicolumn{3}{c}{\textbf{LUTs/ALMs}} &
        \multicolumn{3}{c}{\textbf{Flip flops/Registers}} &
        \multicolumn{3}{c}{\textbf{DSPs}}\\
        \cline{3-5}\cline{6-8}\cline{9-11}
        & &
        \textbf{No opt} & \textbf{Opt} &  &
        \textbf{No opt} & \textbf{Opt} &  &
		\textbf{No opt} & \textbf{Opt} & \\\midrule
        \multirow{2}{*}{cordic}
        & Vivado  & 12629 (23.73\%) & 4791 (9.01\%) & 2.64$\times$ & 20813 (19.56\%) & 5155 (4.84\%) & 4.04$\times$ & 9 (4.09\%) & 9 (4.09\%) & -\\
        & Quartus & 7936 (9.88\%)   & 2903 (3.61\%) & 2.73$\times$ & 14249 (6.48\%)  & 4618 (2.10\%)  & 3.09$\times$ & 5 (3.21\%) & 5 (3.21\%) & -\\\midrule
        \multirow{2}{*}{conv2d}
        & Vivado  & 262659 (493.7\%) & 33517 (63\%)    & 6.42$\times$ & 476794 (448.1\%) & 74224 (69.76\%) & 7.84$\times$ & 16 (7.27\%) & 16 (7.27\%) & -\\
        & Quartus & 155750 (193.8\%) & 19908 (24.78\%) & 7.82$\times$ & 246519 (112.1\%) & 39185 (17.83\%) & 6.29$\times$ & 12 (7.69\%) &12 (7.69\%) & -\\
        \bottomrule
	\end{tabular}
	\caption{Resource utilization and reduction in utilization when using optimisations for synthesized designs.}
    \label{tab:resource_util}
\end{table*}

To assess how well the JuliaHLS toolchain is performing, we examined the number of operations that were present in the generated MLIR for each case study in both the front-end and the \textit{High-to-PreCompile} pipeline in the back-end of the toolchain, and also with and without the MLIR optimisation passes enabled.
Specifically, we captured the following information:
\begin{itemize}
    \item Basic blocks: proxy for control flow complexity
    \item Total operations: proxy for maximum number of hardware units needed (with no resource sharing)
    \item Arithmetic operations: the number of arithmetic operations
    \item Memory copies: amount of data movement
\end{itemize}

The results are shown in Table~\ref{tab:scires_complexity}, where a higher reduction means that the JuliaHLS toolchain has found more opportunities for optimisation.
Overall, the toolchain benefits greatly from including the built-in MLIR optimisations, reducing the number of total operations by nearly half in the cordic example, and reducing the number of memory copies from 87 to 6 in the conv2d example.
These results also demonstrate that the front end optimisations appear to focus more on control flow and integer components, while back end optimisations have the most impact on memory operations.


\subsubsection{Performance analysis}

We next examined the expected performance for the two case studies by building testbenches using Verilator 5.030
We measured the latency for the design as the number of clock cycles between the rising edge of the token handshake / start signal to the rising edge of the result\_valid / done signal, and calculated throughput using the latency assuming a frequency of 100 MHz.
The results for the latency and throughput are in Tables~\ref{tab:performance_analysis:latency}, and~\ref{tab:performance_analysis:throughput}, respectively.
In this testing, Verilator was unable to simulate the unoptimized Verilog generated for the conv2d case study, but did simulate the optimized version, and for the cordic case study, using the optimisations gave a nearly 24$\times$ speedup.

\subsubsection{Synthesis}

We tested the portability of the Verilog for each case study by synthesizing the designs using AMD Vivado 2024.02 and Quartus Prime Pro 2025, and also successfully ran the generated Vivado designs on a Pynq Z1.
The results for each synthesis tool are shown in Table~\ref{tab:resource_util}, showing the resource utilization and the reduction in utilization that enabling optimisations in JuliaHLS provides.
Note that without the JuliaHLS optimisations, the conv2d design cannot fit on either the AMD or altera FPGAs, but with the optimisations it can.
Together, these results clearly show that optimisations implemented by our work are indeed crucial to ensuring that resources are used efficiently and that designs fit on FPGAs. 

\subsection{Comparison to other toolchains}

\begin{table*}[ht!]
	\centering
	\fontsize{9pt}{11pt}\selectfont
    \setlength\extrarowheight{2pt}
	\begin{tabular}{>{\centering\arraybackslash} m{0.05\linewidth}
            >{\centering\arraybackslash} m{0.22\linewidth}|
            >{\centering\arraybackslash} m{0.10\linewidth}
            >{\centering\arraybackslash} m{0.08\linewidth}
            >{\centering\arraybackslash} m{0.05\linewidth}|
            >{\centering\arraybackslash} m{0.07\linewidth}
            >{\centering\arraybackslash} m{0.20\linewidth}}
        \toprule
		\textbf{Design} &
        \textbf{Toolchain} &
        \textbf{LUTS/ALMs} &
        \textbf{Flip flops} &
        \textbf{DSPs} &
        \textbf{Latency (cycles)} &
        \textbf{Maximum throughput\newline (K Op/S)}\\\midrule
		\multirow{2}{*}{implicit else}
        & JuliaHLS (This work) & 2.29\% & 2.52\% & 0\% & 13 & 27548.21\\
        & Dynamatic v2.0       & 1.06\% & 0.63\% & 0\% &  3 & 33333.33\\\midrule
        \multirow{2}{*}{cordic}
        & JuliaHLS (This work) & 9.01\% & 4.83\% & 9\% & 241 & 615.61\\
        & Dynamatic v2.0       & 2.23\% & 0.48\% & 9\% & 97  & 1030.93\\
        \bottomrule
	\end{tabular}
	\caption{Comparison of performance and resource usage between JuliaHLS and Dynamatic v2.0.}
    \label{tab:toolchain_comp}
\end{table*}

To gauge the effectiveness of JuliaHLS against other toolchains, we compare against the MLIR-based dynamically scheduled Dynamatic v2.0 toolchain \cite{JosGue:20}, which compiles C to VHDL using a custom MLIR backend.
We rewrote the `\texttt{implicit\_else}', `\texttt{cordic}' and `\texttt{conv2d\_im2col}' programs in C, and then compiled them with Dynamatic with the default settings.

The `\texttt{implicit\_else}' program was the only program that Dynamatic both compiled correctly and generated a design that satisfied the timing constraints at 100 Mhz.
The `\texttt{cordic}' program failed to meet timing, and the `\texttt{conv2d\_im2col}' program failed to compile entirely as one of the lifting passes crashed the tool when raising from C-style arrays to the `\texttt{affine}' dialect.
After analysis, we believe that the timing violation in `\texttt{cordic}` is because the standard buffer insertion algorithm is too conservative, meaning that Dynamatic's designs are insufficiently pipelined and that performance results only provide a lower bound.

The performance results and resource usage are shown in Table~\ref{tab:toolchain_comp}, showing that compared to JuliaHLS, the Dynamatic designs have lower latency and use fewer resources.
\section{Conclusion}
\label{sec:conclusion}

Overall, the results presented in Section~\ref{sec:results} show that the JuliaHLS toolchain is a viable way of compiling Julia programs into HDL that is both performant and vendor-agnostic, achieving up to 82.6\% of the throughput of ‘state-of-the-art’ toolchains like Dynamatic v2.0 \cite{JosGue:20} while natively supporting linear algebra.

This toolchain is also a step-change from the first Julia HLS toolchain proposed by \citet{BigMcI:22}, compiling not only the simple control flow structures they described but also handling memory-based objects like matrices and vectors in complicated programs like the 2D convolution kernel.
We have also shown the benefits of building an HLS toolchain using an existing framework supported by the HLS community, since using CIRCT allowed us to focus on developing and refining the frontend and initial \textit{High-to-PreCompile} workflows that are unique to Julia instead of having to spend time implementing all the low-level scheduling and generation routines.

As mentioned in Section~\ref{sec:results}, JuliaHLS is currently limited in the designs that it can synthesize using static scheduling due to the currently limited integration of Calyx into the CIRCT framework.
As the Calyx integration with CIRCT becomes more mature, those limitations should be removed and then the JuliaHLS toolchain would be able to compile idiomatic Julia code with either dynamic or static scheduling.


This toolchain has the potential to offer a unique development cycle when paired with Julia’s REPL and JIT compiler --- opening up the possibility of interactive algorithm-accelerator co-development by transparently allowing the same piece of code to be run on the CPU and FPGA.
Consequently, our work can expand the number of users designing and deploying hardware accelerators by making it easier for non-hardware engineers to prototype and deploy them.
Additionally, JuliaHLS can tighten the engineering cycle and can help eliminate the lengthy development process associated with algorithm/accelerator co-design that currently limits widespread adoption of custom hardware accelerators for scientific algorithms \cite{CurFio:23, Kac:25}.

\section{Future Work}
\label{sec:future}

Even though this toolchain captures a large subset of the Julia language and generates functionally `correct' hardware designs, the tool and the ecosystem which it relies upon is still in its infancy.
Producing designs that are competitive against hand-crafted RTL and that outperform mature, industry-backed HLS tools, like AMD Vitis, necessitates continued development and improvement.
Overall, there are three main directions to push the JuliaHLS toolchain forward and build an FPGA ecosystem in Julia: improving the Julia infrastructure, building out new MLIR dialects and transformations, and improving the CIRCT framework and HLS dialects.

Improving the Julia infrastructure takes two forms: working on upstream Julia to improve aspects such as the type inference infrastructure, and building out additional components of the FPGA ecosystem.
Currently, the type inference algorithm, and compiler in general, is very difficult to modify without creating a `custom' Julia version, as shown by \citet{FisSab:18}.
Future work could improve the integration the \texttt{MLIRInterprer} has with the Julia compiler by allowing more fine-grained control over Julia’s ‘middle compiler’ algorithm and the pieces that are reused.

The next step in the JuliaHLS ecosystem would be to allow actually using the compiled accelerators from Julia, not just designing them.
To accomplish this, frameworks such as the Xilinx Runtime (XRT) or Coyote v2 \cite{ramhorstCoyoteV2Raising2025} could be used to wrap the generated IP cores and allow for passing data and control signals between the core and Julia.

Another direction would be to continue developing MLIR dialects relevant to hardware, such as the \texttt{quant} dialect supporting fixed-point arithmetic.
Our work experimented with this approach for fixed point numbers, but there were no open-source passes or transformations, as all existing targets using this dialect for this use-case are closed-source.
Additionally, we found that many scientific and engineering codes rely on mathematical operators in standard libraries, such as trigonometric functions, square root, power, etc.
Developing an ``FPGA standard library'' of these functions or generators for them would enable more applications in Julia to target FPGAs for hardware acceleration.

Finally, the capabilities of the CIRCT library were found to be the main limiting factor in the performance of the hardware, the features that could be included and the ability to perform co-simulation within Julia.
As a result, future work could focus on further development of the `calyx' and `handshake' dialects to produce designs that outperform state-of-the-art designs while remaining fully open-source and integrating with the wider CIRCT ecosystem.
Additionally, the `arcilator' project \cite{ErhSch:23} that compiles CIRCT's hardware dialects into LLVM code (to be run in a JIT) would directly enable gate-level co-simulation from within the Julia REPL and thus integrate the verification process of hardware designs into the toolchain itself.



\clearpage
\bibliographystyle{ACM-Reference-Format}
\bibliography{sample-base}

\appendix

\end{document}